\documentclass[pdflatex,sn-nature,iicol]{sn-jnl}
\usepackage{graphicx}
\usepackage{dcolumn}
\usepackage{bm}
\usepackage{hyperref}
\usepackage{xcolor}
\usepackage{ulem}

\usepackage{flushend}
\usepackage{comment}
\usepackage{cleveref}

\begin{document}
\pagenumbering{Roman}


\title[\hbox{}]{Enhancing European Cooperation in the Search for Dark Matter}

\author[1]{\fnm{Bernard} \sur{Andrieu}}
\author[2]{\fnm{Ties} \sur{Behnke}}
\author[3]{\fnm{Philip} \sur{Bechtle}}
\author[4]{\fnm{Xavier} \sur{Bertou}}
\author[5]{\fnm{Jose} \sur{Busto}}
\author[6]{\fnm{Susana} \sur{Cebrián}}
\author[7]{\fnm{Marco} \sur{Cirelli}}
\author[8]{\fnm{Javier} \sur{De Miguel}}
\author[9]{\fnm{Laurent} \sur{Derome}}
\author[5]{\fnm{Cristinel} \sur{Diaconu}}
\author[10]{\fnm{Caterina} \sur{Doglioni}}
\author[11]{\fnm{Guiliana} \sur{Fiorillo}}
\author[12]{\fnm{Davide} \sur{Franco}}
\author[13]{\fnm{Juan} \sur{Fuster}}
\author[1]{\fnm{Romain} \sur{Gaior}}
\author[14]{\fnm{Erika} \sur{Garutti}}
\author[15]{\fnm{Claudio} \sur{Gatti}}
\author[13]{\fnm{B.} \sur{Gimeno-Martinez}}
\author[1]{\fnm{Fr\'ed\'eric} \sur{Girard}}
\author[16]{\fnm{Roxanne} \sur{Guenette}}
\author[3]{\fnm{Matthias} \sur{Hamer}}
\author[4]{\fnm{Sophie} \sur{Henrot-Versillé}}
\author[4]{\fnm{Thibaut} \sur{Houdy}}
\author[5]{\fnm{Fabrice} \sur{Hubaut}}
\author[13]{\fnm{Adrian} \sur{Irles}}
\author[4]{\fnm{Yoann} \sur{Kermaidic}}
\author[17]{\fnm{Marcin} \sur{Ku\'zniak}}
\author[2]{\fnm{Axel} \sur{Lindner}}
\author[18]{\fnm{Julien} \sur{Masbou}}
\author[15]{\fnm{Giovanni} \sur{Mazzitelli}}
\author[4]{\fnm{Akira} \sur{Miyasaki}}
\author[4]{\fnm{Enrique} \sur{Minaya}}
\author[14,19]{\fnm{Konstantinos} \sur{Nikolopoulos}}
\author[20]{\fnm{Federica} \sur{Petricca}}
\author[4]{\fnm{Roman} \sur{P\"oschl}}
\author[5]{\fnm{Pascal} \sur{Pralavorio}}
\author*[21]{\fnm{Florian} \sur{Reindl}}\email{florian.reindl@tuwien.ac.at}
\author[17,27]{\fnm{Leszek} \sur{Roszkowski}}
\author[8]{\fnm{Daniel} \sur{Santos}}
\author*[21,22]{\fnm{Jochen} \sur{Schieck}}\email{jochen.schieck@oeaw.ac.at}
\author*[2,23]{\fnm{Thomas} \sur{Schörner}}\email{thomas.schoerner@desy.de}
\author[8]{\fnm{Silvia} \sur{Scorza}}
\author[1]{\fnm{Luca} \sur{Scotto~Lavina}}
\author[24]{\fnm{Steinar} \sur{Stapnes}}
\author[4]{\fnm{Achille} \sur{Stocchi}}
\author[25]{\fnm{Maxim} \sur{Titov}}
\author[26]{\fnm{Julia~K.} \sur{Vogel}}
\author[17]{\fnm{Masayuki} \sur{Wada}}
\author[4]{\fnm{Jonathan} \sur{Wilson}}
\author[1]{\fnm{Yajing} \sur{Xing}}
\author*[23]{\fnm{Dirk} 
\sur{Zerwas}}\email{dirk.zerwas@in2p3.fr}

\affil[1]{\orgname{LPNHE, Sorbonne Université, CNRS/IN2P3}, \orgaddress{\city{Paris}, \country{France}}}

\affil[2]{\orgname{Deutsches Elektronen-Synchrotron DESY}, \orgaddress{\city{Hamburg}, \country{Germany}}}

\affil[3]{\orgdiv{Physikalisches Institut}, \orgname{Rheinische Friedrich-Wilhelms Universit\"at Bonn}, \orgaddress{\city{Bonn}, \country{Germany}}}

\affil[4]{\orgname{IJCLab Orsay, Universit\'e Paris-Saclay, CNRS/IN2P3}, \orgaddress{\city{Orsay}, \country{France}}}

\affil[5]{\orgname{CPPM, Aix Marseille Université, CNRS/IN2P3}, \orgaddress{\city{Marseille}, \country{France}}}

\affil[6]{\orgname{CAPA, Universidad de Zaragoza}, \orgaddress{\city{Zaragoza}, \country{Spain}}}

\affil[7]{\orgname{LPTHE, CNRS $\&$ Sorbonne Université}, \orgaddress{\city{Paris}, \country{France}}}

\affil[8]{\orgname{Instituto de Astrofísica de Canarias}, \orgaddress{\city{La Laguna, Tenerife}, \country{Spain}}}

\affil[9]{\orgname{University of Grenoble Alpes, CNRS, Grenoble INP, LPSC-IN2P3}, \orgaddress{\city{Grenoble}, \country{France}}}

\affil[10]{\orgname{University of Manchester}, \orgaddress{\city{Manchester}, \country{UK}}}

\affil[11]{\orgname{Università degli Studi di Napoli Federico II \& INFN}, \orgaddress{\city{Naples}, \country{Italia}}}

\affil[12]{\orgname{APC, Universit\'e de Paris, CNRS/IN2P3}, \orgaddress{\city{Paris}, \country{France}}}%

\affil[13]{\orgname{IFIC, CSIC -- Universitat de Val\`encia}, \orgaddress{\city{Valencia}, \country{Spain}}}

\affil[14]{\orgname{University of Hamburg}, \orgaddress{\city{Hamburg}, \country{Germany}}}

\affil[15]{\orgname{Laboratori Nazionali di Frascati, INFN}, \orgname{\city{Frascati (Roma)}, \country{Italia}}}

\affil[16]{\orgname{University of Manchester}, \orgaddress{\city{Manchester}, \country{UK}}}

\affil[17]{\orgname{AstroCeNT at Nicolaus Copernicus Astronomical Center of the Polish Academy of Sciences}, \orgaddress{\city{Warsaw}, \country{Poland}}}

\affil[18]{\orgname{Subatech, IMT Atlantique, Nantes Universit\'e, CNRS/IN2P3}, \orgaddress{\city{Nantes}, \country{France}}}

\affil[19]{\orgname{University of Birmingham}, \orgaddress{\city{Birmingham}, \country{UK}}}

\affil[20]{\orgname{Max Planck Institute for Physics}, \orgaddress{\city{Garching}, \country{Germany}}}

\affil[21]{\orgname{Atominstitut, Technische Universit\"at Wien}, \orgaddress{\city{Vienna}, \country{Austria}}}

\affil[22]{\orgname{Institut f\"ur Hochenergiephysik der \"Osterreichischen Akademie der Wissenschaften}, \orgaddress{\city{Vienna}, \country{Austria}}}

\affil[23]{\orgname{DMLab, Deutsches Elektronen-Synchrotron DESY}, \orgaddress{\city{Hamburg}, \country{Germany}}}

\affil[24]{\orgname{CERN}, \orgaddress{\city{Geneva}, \country{Switzerland}}}

\affil[25]{\orgname{Institut de recherche sur les lois fondamentales de l'Univers, CEA Saclay}, \orgaddress{\city{Gif sur Yvette}, \country{France}}}

\affil[26]{\orgname{TU Dortmund University}, \orgaddress{\city{Dortmund}, \country{Germany}}}

\affil[27]{\orgname{National Centre for Nuclear Research}, \orgaddress{\city{Otwock}, \country{Poland}}}

\date{\today}

\abstract{The search for dark matter is an exciting topic that is pursued in different communities over a wide range of masses and using a variety of experimental approaches. 
The result is a strongly correlated matrix of activities across Europe and beyond, both on the experimental and the theoretical side. 
We suggest to encourage and foster the collaboration of the involved institutions on technical, scientific and organisational level, in order to realise the synergies that are required to increase the impact of dark matter research and to cope with the increasing experiment sizes. 
The suggested network --- loosely titled "DMInfraNet" --- could be realised as a new initiative of the European strategy or be based on existing structures like iDMEu or DRD. The network can also serve as a nucleus for future joint funding proposals.}

\maketitle
\clearpage
\pagenumbering{arabic}

\setcounter{page}{1}
\section{\label{sec:intro}Introduction}


Elucidating the nature of dark matter is one of the grand challenges of modern physics. Consequently, a whole slew of experimental activities is currently ongoing or being planned in all parts of the world. These activities cover searches of a large variety of theoretically motivated dark matter candidates, ranging over several orders of magnitude in mass and interaction cross-section.  Any successful experimental search for dark matter needs to be accompanied by substantial theory efforts. 

The large variety of dark matter candidates results in very different experimental approaches employing different technologies. Consequently, a lot of very diverse technical expertise exists at various places. Existing and future experiments can significantly benefit from each other and from an improved networking among them. Such interactions can bring the field forward by a  decisive step and are at the heart of the networking idea discussed in this paper.   

    %


The still-unknown nature of dark matter and the resulting variety of experimental approaches make dark matter a prime example for approaches from different angles, i.e.\ from astroparticle, particle and nuclear physics.  This is also reflected in the strategies of the bodies representing the respective communities (APPEC, ECFA and NuPeCC) and in the spirit of the previously established joint ECFA-NuPECC-APPEC symposia (JENAS). 




\section{\label{sec:activities}Dark matter activities and infrastructures in Europe}

The necessary technical effort for dark matter experiments often surpasses the abilities of a single group or institute, frequently requiring significant infrastructures. Among these are e.g.\ underground laboratories, accelerators, substantial cryogenic installations, strong and large-volume magnetic fields, standardised and large-scale read-out and data management systems, etc.

Locally, these infrastructures are normally complemented by dedicated expertise. The facilities and experiments address a whole slew of dark matter candidates at very different mass scales.

The activities are a matrix of labs hosting infrastructure and the actual experiments. While the former is mainly
national or regional, the latter are international. In \cref{app:activities}, we give some (non exhaustive) examples structured by:
\begin{itemize}
\item underground facilities and related infrastructures,
\item accelerator-related facilities,
\item DRD collaborations and potential connections to them, and 
\item concrete experiments. 
\end{itemize}

In addition to these experimental activities, an important theoretical effort is ongoing. This covers building phenomenological dark matter models and studying their consequences, and it also includes the preparation of background predictions (e.g.\ the neutrino fog/floor) as well. The work brings together particle physics questions (particle properties, DM annihilation cross sections) with cosmological considerations. The full range
of dark matter masses from very light, e.g.\ axion-like, to the weak scale and even to primordial black holes is considered.

All the infrastructures and experiments discussed are collaboratively built and exploited by teams from many countries. 

\Cref{sec:europeaffair} gives an overview of activities pursued by different European countries. Most of these activities can also be found in the listing of facilities and experiments in \cref{app:activities} --- altogether, the impression of a strongly correlated matrix of countries, institutions, facilities and experiments arises. 



\section{The motivation}

The 2020 update of the European strategy update~\cite{eppsu2020} emphasised the search for dark matter as one of the big open questions of modern physics~\cite{eppsudelib}:

\begin{itemize}
\item ``The quest for dark matter and the exploration of flavour and fundamental symmetries are crucial components of the search for new physics. [...]
A diverse programme that is complementary to the energy frontier is an essential part of the European particle physics Strategy. Experiments in such diverse areas that offer potential high-impact particle physics programmes at laboratories in Europe should be supported [...]"
\end{itemize}

The 2020 strategy update also emphasises the necessity to strengthen existing and future experiments complementary to those at the energy frontier and to expand cooperation between the relevant research institutions further.


Based on these recommendations and considerations for 2020 and most importantly in light of the tremendous progress and growth of the field, we suggest to better capitalise on existing strengths by enhancing the collaboration between researchers and institutions working on dark matter. The existing, very broad experimental research programme is spread across several European research centres, from underground laboratories to accelerator laboratories. This programme encompasses a large variety of technical, scientific, organisational and financial aspects that provide excellent  opportunities for synergies.

The preparation of future dark matter experiments will require the timely shaping of a strong network of expertise and infrastructure. Such a network would allow the identification and exploitation of synergies and the realisation of mutual support among the involved laboratories and institutions. This is particularly important in view of the fact that the next generation of experiments will usually go beyond the scope of a single laboratory.

For the sake of simplicity, we refer to the ideas for improved cooperation collectively as “DMInfraNet”, which could be realised as a new structure or as an enhancement of an existing one.




\section{The role and tasks}

The main purpose of “DMInfraNet” would be to create a structure for the exchange of technology-related expertise and to facilitate access to the infrastructure for dark matter research. 
This structure should build on existing activities and supplement those that were previously missing: “DMInfraNet” should only address issues that are only partially addressed today and should close gaps.  
This approach will strengthen the expertise in the participating institutions and simplify the transfer of technology and knowledge. This also includes all theoretical work carried out in the participating institutions.  

The aim of this network should be to

\begin{itemize}
\item act as a supportive network of related activities in the field of dark matter searches, with focusing on mutual support for existing and future projects; 
\item identify topics of common interest, bundle existing knowledge and interests, and to act as a sounding board for the exchange of ideas, needs, and expertise in the context of the realisation and operation of dark matter experiments;
\item facilitate access to various infrastructures and strengthen existing competence centres or support the establishment of new ones;   
\item help the community to speak with one voice and to prepare grounds for joint activities, in particular joint funding applications;  
\item set up the necessary communication structures and collaborative tools;
\item providing a common framework for solutions to common questions such as a software eco-system;
\item implement a common approach to computing, data storage and data access issues, also in view of increasing open and FAIR data requirements. 
\end{itemize}


In addition to the scientific and organisational aspects highlighted above, we would also like to point out educational opportunities provided by smaller-scale particle physics experiments. Their much shorter timescales and more lightweight formal structures typically allow early-career scientists and technical personnel to experience different phases of an experimental endeavor within just a few years, to acquire project management skills and to learn to take responsibilities before being confronted with similar challenges in large collaborations working on collider experiments. 

Finally, on the funding side, it should be explored whether the intended network could also prepare grounds for common future funding applications on the European level with the aim of setting up new experiments and necessary infrastructures for their development and construction at the various institute locations.


\section{Implementation}


The network is supposed to span all theory models, technologies, experiments, institutions, and countries involved in dark matter research. It particularly aims at providing communication, exchange and advice between the partners, at facilitating access to and use of existing infrastructures and technologies, and at finding and jointly implementing solutions to common problems and issues in the field.    

Beyond mutual technical, scientific and organisational support between the partners, the network could also be geared towards strengthening the strategic voice of dark matter research in Europe, towards preparing common funding applications and, ultimately, towards the official recognition as a CERN activity. 

Given these considerations, three possible options for the proposed structure have been identified: 

\begin{itemize}

\item "DMInfraNet" could thus be a new structure established by an encouragement of the ESPPU process.

\item "DMInfraNet" could also be an extension of the existing DRD activities, which were established as a result of the last strategy update in 2020. While the successfully founded DRDs focus on detector technology development, this new part of the programme could focus on implementing existing and new technologies. The existing DRD activities cover the technology readiness level (TRL)~\cite{NASA_TRL} 1-6, while this new initiative targets, among others, the remaining levels 7-9, system and subsystem development, system tests, launch and operation. This would supplement the existing DRD programme, leading to full coverage of all TRLs. 

\item Another implementation could be an enhancement of the iDMEu initiative: iDMEu~\cite{Cirelli:2023gko} is a collective
effort by a group of particle and astroparticle physicists aiming to set up an online resource
meta-repository, a common discussion platform and a series of meetings on everything concerning dark matter.

 \end{itemize}

 Europe, and in particular CERN, should play an important role in the search for dark matter. CERN support for relevant activities, with adequate resources allocated,   would be very timely and fully in line with previous strategy considerations.

\clearpage 

\appendix
\newpage

\section{\label{app:activities}Dark matter activities and infrastructures in Europe}



As described above, the activities in dark matter research form a matrix of laboratories hosting infrastructure and the actual experiments. While the former are mainly
national or regional, the latter are mostly international. We therefore subdivide the following section into underground facilities,
accelerator facilities, and experiments. All entries are sorted alphabetically as much as possible, according to their acronyms. 

\subsection{Underground facilities and related infrastructures}

Underground facilities play a crucial role in low-background DM searches. The main facilities in Europe are:
\begin{itemize}
    \item Boulby in the UK is a dark matter research facility and provider of facilities for large-scale material tests.

    \item The Laboratori Nazionali del Gran Sasso (LNGS) in Italy is the largest underground laboratory in Europe devoted to neutrino and astroparticle physics.

    \item The Laboratorio Subterráneo de Canfranc (LSC) in Spain is located 800~m deep below the Pyrenees at Canfranc, the main research areas are dark matter, neutrinos and also biology.
        
    \item The Laboratoire Souterrain de Modane (LSM) in France is the deepest underground Laboratory in Europe. It is dedicated to the development of  astroparticle and nuclear physics programmes.

\end{itemize}

In addition to these large facilities, infrastructures and expertise is available in many labs:
\begin{itemize}
    \item University of Zaragoza: At the Center for Astroparticles and High Energy Physics (CAPA, there are laboratories for R\&D activities related to WIMP and axion searches (using mainly scintillators and micromegas readouts) and to low-radiaoctivity techniques.

    \item CPPM Marseille: The lab hosts a radon platform the purpose of which is to study the main problems associated with radon-induced background noise in low-energy neutrino physics and in direct searches for dark matter. The background noise induced by radon and its progeny is very often the most difficult component to eliminate and the ultimate limitation for a large number of experiments. The goal is to achieve a filtration quality of the order of $\mu$Bq/m$^3$.

    \item IJCLab  is hosting the Cryogenic Quantum Detectors (DQC) platform dedicated to the design, fabrication and calibration of cryogenic (sub-K) solid-state detectors. 
    
    \item LPNHE Paris is hosting and running XeLab, a cryogenic dual-phase Xe TPC test platform.

    \item LPSC-LSM is hosting and running the D2S2 platform for the "Directional Detection of dark matter and neutrons for Science and Society", with a special focus on ionisation quenching factor measurements.

    \item University Federico II Napoli is hosting and running CryoLab, a cryogenic dual-phase Ar TPC test platform.
\end{itemize}

\subsection{Accelerator-related facilities}

Prominent examples for accelerator facilities providing beams for dark matter searches are: 
\begin{itemize}
    \item At CERN~\cite{SHiP:2021nfo}, the SPS provides the beams for fixed-target / beam dump experiments.
    
    \item At DESY, the ELBEX beamline at the European XFEL will provide high rates of 17 GeV electrons.

    \item ELSA~\cite{Bechtle:2024atq} provides an intense electron beam of 3.5~GeV for the Lohengrin experiment. 

    \item At LNF, the Beam Test Facility (BTF) provides an intense beam of positrons, e.g.\ for the positron beam from the DA$\Phi$NE Linac to the PADME experiment.
\end{itemize}

Accelerator-related infrastructures are magnets and RF cavities as well as cryogenics and mechanical platforms. 
RF cavities are part of the signal detection chain for dark matter experiments. Here
are some examples:
\begin{itemize}

    \item CEA holds leading expertise in magnet design and related issues.

    \item At CPPM Marseille, precision mechanics from pixel detector developments are applied to dark axion searches.

    \item DESY in Hamburg provides a 1.2~T magnet for the test beam (PCMAG), and it is working on a 9~T magnet. DESY also plans a versatile magnet infrastructure on a movable platform. The cryo-platform at DESY, a versatile cryogenic facility, provides helium at 4~K to up to three experiments in the HERA north area. DESY's expertise in RF, quantum sensing, extremely low-noise detector systems and high-precision optical interferometry is essential for signal detection.

    \item The IFIC High-Gradient (HG) Radio-Frequency (RF) laboratory in Valencia hosts a high-power infrastructure for testing HG S-band normal-conducting RF accelerating structures to the study of HG phenomena. In addition, IFIC will be the host of a proton-ion accelerator. The construction of this accelerator has started.
    
    \item IJCLab in Orsay holds important  expertise on RF and its cryo-platform. In particular, the ALTO facility at  provides directional, pulsed neutron beams, fast (0.5-4.0 MeV) or epithermal (50-200 keV), using its 15~MV tandem accelerator. These beams are particularly well-suited for characterising the response of noble liquids for WIMP searches near detection thresholds.
        
    \item LNF (Frascati) holds infrastructure for axion searches and hosts the QUAX$_{a\gamma}$ and PADME experiments.
    
    \item LNL (Legnaro) provides the infrastructure for axion searches hosting the QUAX$_{a\gamma}$, QUAX$_{ae}$ and QUAX$_{gpgs}$ experiments.
    
    \item Mainz hosts the MAMI accelerator for electrons of up to 1.6~GeV. MAMI mostly serves experiments inquiring into the structure of hadrons, but   allows e.g.\ for tests of nuclear models important for neutrino experiments, and for other other BSM-related measurements.

\end{itemize}

\subsection{\label{sec:drds}The DRD collaborations}

The DRD collaborations, created as a follow-up of the previous strategy update, are of great importance for collaborative detector development for dark matter searches:
The gaseous detectors of DRD1 are relevant for direct searches (TPCs), as is DRD2 with liquid detectors.   Indirect searches rely heavily on the developments of 
semi-conductors in DRD4 and calorimetry in DRD6. Transverse to both indirect and direct searches is the development of highly performant electronics in DRD7.
Innovative approaches are pursued in DRD5 on quantum sensors that enable sensitivity to sub-eV energy deposits from dark matter interactions.

\subsection{Experiments}

The experiments in the field of dark matter searches can be subdivided into different categories depending on the type of dark matter particle
they are searching for. For simplicity, no separation is made between experiments ongoing and in preparation.

\paragraph{WIMP searches using liquefied noble elements such as Argon and Xenon:}
\begin{itemize}

    \item ARGO: a multi-100-ton experiments to reach the neutrino floor/fog at SNOLAB.

    \item DarkSide-20k~\cite{DarkSide-20k:2017zyg}: based on a dual-phase argon time projection chamber (TPC), under construction at LNGS, with the objective to achieve a 90\% exclusion sensitivity for WIMP-nucleon cross-sections O($10^{-48} \textrm{cm}^{2}$) at 100 GeV WIMP mass scale. Using only the ionization channel, it also has sensitivity at a few GeV WIMPs. 

    \item GADMC (Global Argon Dark Matter Collaboration)~\cite{DarkSide-20k:2017zyg}: collaboration of the DEAP, DarkSide, CLEAN and ArDM experiments to build a detector with a liquid argon mass above 20 tonnes (DarkSide-20k) at the LNGS. 
    
    \item LUX-ZEPLIN~\cite{LZ:2019sgr}: direct detection search for cosmic WIMP dark matter particles with a large liquid Xenon time projection chamber installed at SURF.

    \item XENONnT~\cite{xenoncollaboration2024xenonntdarkmatterexperiment} utilises a liquid Xenon detector to search for signals of interactions between WIMPs and atomic nuclei, playing a leading role in the WIMP scenario search.
    
   \item XLZD~\cite{Aalbers:2022dzr,DARWIN:2016hyl}: collaboration of XENON, LUX-ZEPLIN and Darwin to prepare the next generation Xenon experiment.

\end{itemize}

\paragraph{NaI-based experiments:}
\begin{itemize}

\item ANAIS~\cite{anais_2024} is based on about 100~kg of NaI(Tl) scintillators operated at LSC to measure WIMPs, in particular the annual modulation expected from the Earth's motion around the Sun. The current results are inconsistent with DAMA-LIBRA at a 4$\sigma$ confidence level, and data taking will continue until reaching a 5$\sigma$ sensitivity. 

\item COSINE~\cite{PhysRevD.106.052005}: using also about 100~kg of NaI(Tl) detectors with the aim to explore the DAMA/LIBRA annual modulation signal, operated in South Korea. Published results are consistent with both the modulation amplitude reported by DAMA/LIBRA and the no-modulation case. 

\item COSINUS~\cite{COSINUS:2023kqd}: based on cryogenic scintillating calorimeters with undoped sodium iodide (NaI) crystals, providing an independent verification of the DAMA/LIBRA results with a different detection technology.

\item DAMA/LIBRA~\cite{DAMA:2008bis}: already finished, used thallium-doped sodium iodide crystals (250~kg) at LNGS to detect annual variations in the interaction rate of WIMPs with ordinary matter, claiming the observation of an annually modulated signal that could indicate the presence of WIMPs.

\item SABRE~\cite{SABRE:2018lfp} utilises highly pure sodium iodide (NaI) crystals to detect dark matter, confirming or refuting the results of the DAMA/LIBRA experiment by comparing results from the southern hemisphere.

\end{itemize}

\paragraph{Experiments aiming primarily at lower masses:}
\begin{itemize}

\item CRESST~\cite{CRESST-II:2014ezs}: for sub-GeV dark matter search, uses cryogenic detectors to search for dark matter signals through the interaction of WIMPs with the nuclei of target materials at extremely low temperatures, with improved techniques for discriminating beta/gamma background.

\item DELight~\cite{10.21468/SciPostPhysProc.12.016} is being designed for light dark matter searches down to masses well below 100~MeV. It will use superfluid helium, ideal to search for dark matter nucleus scattering given the light target nucleus and the potential for discriminating against beta/gamma backgrounds using the interaction dependent partitioning of energy into photons and quasiparticles~\cite{PhysRevD.111.032013}.

\item DAMIC-M~\cite{DAMIC:Arnquist_2023} employs thick, ultra-low noise, fully depleted silicon charged-coupled devices (CCDs) to search for dark matter particles. A novel skipper readout implemented in the CCDs provides single electron resolution through multiple non-destructive measurements of the individual pixel charge, pushing the detection threshold to the eV scale. DAMIC-M installed underground at LSM is searching for low mass WIMPs ($<$5GeV/c$^{2})$ through the nuclear recoil they would induce on silicon nuclei. The experiment  will also probe much lighter candidate such as particles of the hidden sector through the electronic recoil they would provoke.

\item NEWS-G/DarkSPHERE~\cite{NEWS-G:2023qwh}: searches for light dark matter in the sub-GeV (0.05--10~GeV) mass region with spherical proportional counters filled with light gas mixtures~\cite{Knights:2025ogz}.The NEWS-G collaboration obtained results with a 1.4~m diameter sphere operating at LSM~\cite{NEWS-G:2024jms}. Currently, the detector collects data at SNOLAB. The planned DarkSPHERE detector will be fully electroformed underground at Boulby and could provide sensitivity down to the neutrino floor~\cite{NEWS-G:2023qwh}. 

\item The TESSERACT~\cite{Billard:2024zvc} (Transition Edge Sensors with Sub-eV Resolution And Cryogenic Targets) experiment aims at detecting light dark matter via its interaction with ultra-sensitive new-generation cryogenic detectors. TESSERACT aims to probe dark matter over 12 orders of magnitude in mass.
        
\item TREX-DM~\cite{Castel:2023pby} searches for WIMPs with a low background chamber installed at the LSC with micromegas operated with different Argon and Neon mixtures. Designed to detect WIMPs with massed of the order of 10~GeV, its sensitivity region is being extended to lighter masses below 1~GeV.
    
\end{itemize}

\paragraph{Developments for directional detection, aiming at increasing the sensitivity in the neutrino fog/floor:}
\begin{itemize}

\item CYGNO~\cite{Amaro:2022gub} employs a high-resolution gaseous time projection chamber (TPC) with optical readout using scientific CMOS (sCMOS) sensors, aiming to detect WIMPs with masses above 1~GeV, exploiting the directionality capability offered by high granularity readout.

\item CYGNUS~\cite{Vahsen:2020pzb} aims to exploit multiple TPC operated at atmosopheric pressure to measure the direction of the nuclear recoil in WIMP-nucleous interactions to improve the sensitivity for dark matter searches when reaching the neutrino floor/fog.

\item DRIFT~\cite{Daw:2011wq}, the experiment operated at Boulby, is based on a low-pressure gaseous chamber enabling detection of WIMP dark matter through directional measurements.

\item MIMAC~\cite{Iguaz:2011yc}: The MIMAC (MIcro-tpc MAtrix of Chambers) collaboration has developed, in the last years, an original prototype detector based on the direct coupling of a large pixelated micromegas with a special developed fast self-triggered electronics, showing the feasibility of a new generation of directional detectors.

\end{itemize}

\paragraph{Searches for axions and axion like particles:}
\begin{itemize}

\item ALPS II~\cite{alps2}: the second generation of the Any Light Particle Search experiment ALPS II has been in operation since 2024. It is a light-shining-through-a-wall-experiment looking for axions and similar particles in the sub-milli-eV regime.

\item (Baby)IAXO~\cite{babyiaxo_cdr}: IAXO is a helioscope designed to search for axions and axion-like particles produced in the sun, for which BabyIAXO is scaled version conceived to test all IAXO subsystems. BabyIAXO is expected to be sensitive to axion-photon couplings down to $1.5\cdot 10^{-11}\mathrm{GeV}^{-1}$, and masses up to $0.25\mathrm{eV}$

\item CADEx~\cite{Aja:2022csb}: searches for axions at the LSC using a microwave resonant cavity haloscope in a high static magnetic field coupled to a highly sensitive detecting system. It is sensitive to masses of several hundred $\mu\mathrm{eV}$.

\item DALI is a new-generation haloscope based on a magnetized phased-array (MPA) concept. It is designed to achieve QCD-level sensitivity to axion dark matter in the 25--400 $\mu$eV range ($\sim$6--100 GHz), while also enabling ultra-sensitive searches for dark-photon dark matter and high-frequency gravitational waves. The MPA haloscope, to be hosted at the Instituto de Astrofísica de Canarias (IAC) in Spain, is currently in the prototyping phase ~\cite{DeMiguel2021, DeMiguel2023nmz, Cabrera2023qkt, 2024JInst19P1022H, Hernandez-Cabrera:2024oek}.

\item Lohengrin~\cite{Bechtle:2024atq}: a light dark matter search experiment at ELSA based on the fixed-target missing momentum based technique for searching for dark-sector particles.

\item LUXE(-NPOD)~\cite{LUXE:2023crk} will reach and go beyond the Schwinger limit at high energy and high intensity. It will also perform searches for axion-like dark matter particles using the accelerator as a beam dump.

\item MADMAX~\cite{Garcia:2024xzc} ( MAgnetized Disk and Mirror Axion eXperiment) employs a dielectric haloscope with precisely positioned sapphire disks and a mirror to resonantly enhance the axion-induced microwave signal in a strong magnetic field. The sensitivity is in the O(100) $\mu$eV mass range.

\item NuSTAR~\cite{Ruz:2024gkl}: the Nuclear Spectroscopic Telescope Array leverages the sun's magnetic field for the conversion of axions and ALPs into X-rays. The search is sensitive to the mass region of $10^{-4}$ to $10^{-7}$~eV. 

\item PADME~\cite{Raggi:2014zpa} is devoted to the search for any new light particle in the mass range 2~MeV~$\leq M\leq 23$~MeV. While originally focusing on probing various dark photon models, the chosen experimental technique is suitable also for addressing ALPs and light dark Higgs boson scenarios. 

\item QUAX$_{a\gamma}$~\cite{QUAX:2023gop,QUAX:2024fut}: search for axion dark-matter through the axion-photon coupling with two Sikivie's haloscopes in a mass range around 40~$\mu$eV corresponding to frequencies of operation of about 10~GHz. Two haloscopes, located at LNL and LNF, are both composed by a resonant cavity surrounded by a superconducting solenoid-magnet and inserted in a dilution refrigerator.

\item QUAX$_{ae}$~\cite{QUAX:2020adt}: a ferromagnetic haloscope searching for axion dark matter.

\item QUAX$_{gpgs}$~\cite{Crescini:2016lwj}: search for a 5th force induced by the monopole-dipole coupling mediated by a pseudoscalar boson.

\item RADES~\cite{Melcon:2018dba} (Relic Axion Detector Exploratory Setup)  is an exploratory project aiming to search dark matter axions in microwave and millimetre frequency ranges. RADES scientists developed a general theoretical framework for the analysis of haloscopes, which was applied to the design of different configurations implemented in rectangular waveguide technology to be operated in different experiments, such as the CAST experiment of CERN or IAXO/BabyIAXO at DESY.

\item RadioAxion~\cite{Broggini:2024udi} aims to detect axion dark matter by observing time-modulated changes in the decay constants of Americium-241 (alpha decay) and Potassium-40 (electron-capture decay).

\item SHiP~\cite{SHiP:2021nfo}: designed to search for feebly interacting particles, search for light dark matter through recoil signatures.

\end{itemize}


In addition to the direct and acceleration searches listed above, there is a wide experimental effort in indirect searches, i.e.\ trying to detect the signals of the annihilation or decays of DM particles in astrophysical environments. These signals include charged cosmic rays, photons and neutrinos. Among the current and imminent experiments mostly devoted to charged cosmis rays, we mention AMS-02~\cite{Battiston:2008zza}, CALET~\cite{CALET:2018bqg}, DAMPE~\cite{DAMPE:2017cev} and GAPS~\cite{GAPS:2013gli}. 
Among those dedicated to low-energy photons we list
COSI~\cite{Tomsick:2023aue}, INTEGRAL~\cite{Winkler:2003nn} and XMM-Newton~\cite{XMM:2001haf,Struder:2001bh,Turner:2000jy}. 
A few experiments focus on high-energy photons --  
CTA~\cite{2011ExA....32..193A},
FERMI~\cite{Fermi-LAT:2009ihh},
HESS~\cite{Bernlohr:2003tfz,Cornils:2003ve,HESS:2006fka}, MAGIC~\cite{2016APh....72...76A} and VERITAS~\cite{Weekes:2001pd}  -- or 
very high-energy photons --  HAWC~\cite{HAWC:2015dxn} and LHAASO~\cite{LHAASO:2019qtb}. 
Finally, some target neutrinos: 
GRAND~\cite{GRAND:2018iaj},
ICECUBE~\cite{Halzen:2010yj}, 
KM3NET~\cite{LAHMANN20121209} and 
Super-KAMIOKANDE~\cite{Super-Kamiokande:2002weg}.

\section{\label{sec:europeaffair}Dark matter: a European affaire}

The infrastructures and experiments discussed above are collaboratively built and exploited by teams from many different countries. 
Some examples (in alphabetical order) are given in the following. This results in a strongly correlated matrix.

\textbf{Austria} is involved in the direct dark matter detection experiments CRESST and COSINUS hosted at LNGS. The MaglevHunt experiment, currently under construction and to be operated in Austria, uses levitating mechanical sensors to detect DM.

\textbf{France} For WIMP searches, the IN2P3 labs LPNHE Paris and Subatech Nantes are involved in the XENON experiment at LNGS
as well as being members of the Darwin and XLZD collaborations. Both labs and IJCLab participate in the
DAMIC-M experiment, which is being installed at LSM. 
The Astroparticle and Cosmology laboratory (APC) in Paris has a leading role in direct dark matter searches in the DarkSide-20k experiment in which the CPPM Marseille is involved as well.

CPPM Marseille and IJCLab are participating in the axion searches in the MADMAX experiment. 
Saclay is part of the BabyIAXO experiment. 

The Tesseract project is carried by the IP2I Lyon, LPSC Grenoble and IJCLab

The expertise on highly granular calorimeters of IJCLab has led to participation in LUXE-NPOD as well as in 
contributions to the study of the Lohengrin detector and to studies of a participation in SHiP.

At IJCLab, the theory group concentrates on the production of dark matter in the early Universe, in non-standard cosmology.
The interplay between inflation and reheating is studied, combining constraints from direct detection, indirect detection
 and reheating temperature. The group is also specialized in Higgs or Z'-type portal models, where the search for Z', or the physics of the Higgs is directly connected
to the direct-detection cross section.

Some theory efforts, on WIMPs, sub-GeV DM, Primordial Black Holes and other candidates, are delevoped at the LPTHE.

\textbf{Germany} Dark matter searches in Germany have strong contributions from the Helmholtz centres DESY and KIT and also numerous activities at universities. At DESY, the international axion search experiments ALPS II, BabyIAXO and MADMAX are operated / prepared, as is the strong-field QED and new physics search experiment LUXE-NPOD. At KIT and at the Max Planck Institute for Nuclear Physics (MPI-K), the Darwin/XLZD experiment is pursued; MPI-K is also involved in XENON. The DELight experiment us pursued by KIT and the universities in Heidelberg and Freiburg. Further university activities comprise the ELSA machine and Lohengrin experiment (Bonn), activities at MAMI in Mainz, and also in Mainz and elsewhere the GNOME network of magnetometers for exotic fields. The NuSTAR experiment is pursued at the University of Dortmund, which also plays an important role in BabyIAXO.

 \textbf{France$+$Germany} created the CNRS-Helmholtz Dark Matter Lab (DMLab) carried by the CNRS, DESY, KIT and GSI to enhance bilateral collaboration on all aspects of dark matter, from detectors to theory.

\textbf{Italy} The activity for searching WIMP dark matter in Italy is mainly centered on one of the largest and most important infrastructures in the world: the National Laboratory of Gran Sasso (LNGS) of INFN. This facility, in collaboration with other worldwide and INFN laboratories, hosts very important experiments for WIMP and sub-GeV dark matter searches.

Moreover, LNGS and INFN laboratories host many R\&D projects (e.g.\ Nucleus, BULLKID, etc.) and collaborate worldwide on the dark matter search strategy~\cite{Billard_2022}, participating in the most important underground, cosmic rays, and space experiments, searching for direct and indirect dark matter candidates.

The search for light dark matter candidates such as dark photons and axions is concentrated at the INFN National Laboratories of Frascati (LNF) and Legnaro (LNL), with some activity on this topic at LNGS and at EGO-VIRGO. 

Frascati, explicitly mentioned in the previous European Strategy, has put together a program for dark matter search~\cite{Gatti:2021cel} that includes axion searches leveraging expertise, e.g.\ on RF cavities, and dark photon searches with a fixed-target program of their beam test facility. 
This activity is carried out by the two experiments PADME and QUAX@LNF.
PADME exploits the positron beam at LNF

The QUAX$_{a\gamma}$ experiment has two haloscopes, located at LNL~\cite{QUAX:2023gop} and LNF~\cite{QUAX:2024fut}. A second haloscope will be build at LNF recycling a 3~m bore NbTi magnet for the FLASH experiment~\cite{FLASH}. FLASH will search for high frequency gravitational waves and axions at a frequency around 100 MHz.

At LNL axions are searched through the axion-electron coupling with the QUAX$_{ae}$ experiment~\cite{QUAX:2020adt} and with the QUAX$_{gpgs}$ experiment~\cite{Crescini:2016lwj}.

At LNGS, the RadioAxion experiment~\cite{Broggini:2024udi} aims to detect axion dark matter.

Finally, dark photon dark matter that could couple to gravitational wave interferometers was searched using data from Advanced LIGO and Virgo~\cite{LIGOScientific:2021ffg}.


\textbf{Spain} For the direct detection of WIMPs, and in close connection with LSC, CAPA from the University of Zaragoza is mainly involved in ANAIS, DarkSide-20k and TREX-DM. CIEMAT (``Centro de Investigaciones Energéticas, Medioambientales y Tecnológicas'') at Madrid collaborates in the DarkSide-20k and DEAP experiments, and IFCA (``Instituto de Física de Cantabria'') in DAMIC-M. Concerning axion experiments, there are CADEx at Canfranc and DALI at IAC (Canarias), and there are involvements in RADES. Additionally, the University of Zaragoza, IFIC Valencia and other groups are contributing to BabyIAXO for axion searches.

IFIC is also involved in the preparation of LUXE, LUXE-NPOD and Lohengrin through its expertise on the development of highly granular electromagnetic calorimeters for collider experiments. IFIC is also contributing to the ELBEX project at DESY.


\textbf{Poland} 
Jagiellonian University in Krakow and Astrocent in Warsaw are involved in DEAP-3600 and DarkSide-20k.

\bibliographystyle{naturemag}
\bibliography{DMInfraNet}


\end{document}